\documentclass[aip,pop,reprint]{revtex4-1}

\usepackage{graphicx}
\usepackage{verbatim}
\usepackage{amsmath}
\usepackage{tabularx}
\usepackage{titlesec}
\usepackage[loading]{tracefnt}
\newcommand{\pd}[2]{\frac{\partial #1}{\partial #2}}
\newcommand{\gyav}[1]{\left\langle #1 \right\rangle_{\mathbf{R}_\alpha}}
\newcommand{\vdrift}[2][s]{\ensuremath{\mathbf{v_{d}}_{s}}}
\usepackage{natbib}

\begin{document}
\title[]{Transport and deceleration of fusion products in microturbulence}

\author{George J. Wilkie$^{1,2}$,  Ian G. Abel$^{3}$, Matt Landreman$^{1}$, William Dorland$^{1}$}

\address{$^1$University of Maryland; IREAP; A.V. Williams Building; College Park, MD 20742 \\ $^2$Chalmers University; Department of Physics; Fysikg\aa rden 1; 41454 Gothenburg, Sweden \\ $^3$Princeton University, PCTS, 410 Jadwin Hall, Princeton, NJ 08544}

\begin{abstract}

The velocity-space distribution of alpha particles born in fusion devices is subject to modification at moderate energies due to turbulent transport. Therefore, one must calculate the evolution of an equilibrium distribution whose functional form is not known \emph{a priori}. Using a novel technique, applicable to any trace impurity, we have made this calculation not only possible, but particularly efficient. We demonstrate a microturbulence-induced departure from the local slowing-down distribution, an inversion of the energy distribution, and associated modifications to the alpha heating and pressure profiles in an ITER-like scenario.
\end{abstract}

\maketitle

\section{Introduction}

Alpha particles are relied upon for heating a burning deuterium-tritium (DT) fusion reactor~\cite{ITERPhysics1999}. The distribution of alpha particles in both radius and energy is therefore of critical importance for fusion. In addition to other important effects such as Alfv\'en eigenmodes~\cite{Fu1989,Breizman1995a,Gunter2007}, the ubiquitous microturbulence provides a background level of transport which has the potential to affect the radial and energy distribution of alpha particles. It is typically assumed that alpha particles slow-down via collisions locally on a flux surface, but recent results~\cite{Wilkie2015} suggest that this could be violated at moderate energies. In this case, the distribution function is not known \emph{a priori} when performing turbulence simulations, and solving for the transport of a general non-Maxwellian distribution function would therefore be a computationally monumental task. Here, we assume alpha particles \emph{passively} respond to the turbulence, an approximation that is widely used in this context~\cite{Estrada-Mila2006,Angioni2008,Hauff2009,Pueschel2012}. Recently-discovered effects~\cite{Citrin2013} close to the kinetic ballooning threshold notwithstanding, this trace approximation is generally a good one for alpha particles in microturbulence which is driven by primarily electrostatic modes~\cite{Angioni2008,Wilkie2015}.

In the low-collisionality gyrokinetic hierarchy, the transport equation averaged over pitch angle reads~\cite{Waltz2013,Wilkie2015a}
\begin{equation}  \label{transporteqn}
\frac{\partial F_{0s} }{\partial t} + \frac{1}{V'} \frac{\partial}{\partial r} V' \Gamma_r + \frac{1}{v^2} \pd{}{v} v^2 \Gamma_v =  S_\alpha,
\end{equation}
where $V=V(r)$ is the volume enclosed by the flux surface labelled by $r$, the half-width at the height of the magnetic axis. $F_{0\alpha}$ is the slowly-varying distribution of alpha particles in the $r-v$ phase space. The energy-dependent source of alpha particles $S_\alpha$ is well-approximated by~\cite{Brysk1973,Appelbe2011} $S_\alpha~\propto~\exp \left[ - 5 m_\alpha^2 \left( v^2 - v_\alpha^2 \right) / 64 T_i E_\alpha \right] $, and has an overall magnitude so that the total source is that given by \citep{Huba2004}. The alpha particle mass is $m_\alpha$, the alpha birth energy is $E_\alpha  = 3.5 \mathrm{MeV} = m_\alpha v_\alpha^2 / 2$, and $T_i$ is the temperature of the reactant ions. The radial flux due to turbulence is defined as:
\begin{equation} \label{rfluxdef}
\Gamma_r \equiv \left\langle \sum\limits_{\sigma_\|} \int  h_\alpha  \mathbf{v_\chi} \cdot\nabla r \, \frac{\pi B \mathrm{d} \lambda }{ \sqrt{1 - \lambda B}} \right\rangle_{t,\psi},
\end{equation}
where the pitch angle coordinate $\lambda \equiv E / \mu$ is the ratio between the energy and the magnetic moment, and $\sigma_\|$ is the sign of the velocity parallel to the magnetic field (which has magnitude $B$ and points in the direction of the unit vector $\mathbf{b}$). The non-adiabatic part of the fluctuating alpha particle distribution is $h_\alpha$, and $\mathbf{v}_\chi \equiv \frac{c}{B}\mathbf{b} \times \nabla \gyav{\chi}$ characterizes the drift due to the turbulent electromagnetic potential $\chi \equiv \phi - \frac{v_\|}{c} A_\|$ (with $\phi$ and $A_\|$ respectively the electrostatic and the parallel component of electromagnetic potentials). The notation $\left\langle \ldots \right\rangle_{t,\psi}$ signifies a time-average over many decorrelation times and a spatial average over a flux tube, and $\gyav{}$ is the gyroaverage at fixed gyrocenter $\mathbf{R}_\alpha$. The flux in velocity space is defined as:
\begin{align} 
\Gamma_v \equiv& -v \nu_s F_{0\alpha} - \frac{1}{2} v^2 \nu_\| \pd{F_{0\alpha}}{v}  \\
&+ Z_\alpha e \left\langle \sum\limits_{\sigma_\|} \int  h_\alpha  \gyav{\pd{\chi}{t}}  \, \frac{\pi B \mathrm{d} \lambda }{ \sqrt{1 - \lambda B}} \right\rangle_{t,\psi}, \nonumber
\label{vfluxdef}
\end{align}
and includes both the test-particle energy scattering from the collision operator (with $\nu_s$ and $\nu_\|$ are defined in \citep{Helander2002a}, and summed over all bulk species), and the turbulent heating/cooling of alpha particles.

The fluxes $\Gamma_r$ and $\Gamma_v$ are both expected to decrease rapidly with energy due to the large magnetic drift orbits and Larmor orbits of high-energy alpha particles \cite{Hauff2009,Zhang2010,Pueschel2012}. Therefore, high-energy alpha particles are expected to be well-confined with respect to microturbulence, while cooled-down helium in thermal equilibrium with the bulk plasma (\emph{i.e.}, ``ash'') transports similarly to the ions. At what energy this transition can be expected to occur and what the consequences are for alpha particle physics is the subject of this letter.

\section{The \textsc{T3CORE} transport code}

The equilibrium bulk plasma density and temperature can be evolved using turbulent fluxes computed by local flux-tube codes, an approach valid in the $\rho^* \rightarrow 0$ multiscale limit~\cite{Parra2015}. Tools such as \textsc{TRINITY}~\cite{Barnes2010} and \textsc{TGYRO}~\cite{Candy2009} have been developed to this end, and have successfully recreated the experimental profiles of plasma density and temperature. However, these simulations are expensive (requiring about one million core hours for a single case) because of the need to repeatedly run turbulence simulations to steady-state. Global simulations that do not take advantage of the separation of scales are even more expensive. 

In gyrokinetic simulations, the evolution of a trace species is generally as expensive as the bulk ions or electrons, which would make a transport calculation that much more expensive. Furthermore, if one does not know the form of the equilibrium distribution (as is the case for fast particles like alphas), an entire grid of $F_{0\alpha}(v)$ at each flux surface is necessary to specify and adjust, as opposed to a small set of parameters (e.g. $n$ and $T$ for a Maxwellian). Therefore, if one were to include a trace non-Maxwellian species in a \textsc{Trinity}-like simulation, it would become at least a further order of magnitude more expensive. This section outlines a method which improves upon this existing state-of-the-art by several orders of magnitude by taking advantage of the trace approximation. 
%Simplifying assumptions such as gyro-fluid closures and ad-hoc turbulence models are therefore useful and appropriate. Here we present an additional one that relies upon the trace approximation. 

One can schematically write the gyrokinetic equation for alpha particles in the form:
\begin{equation} \label{schematicGK}
\mathcal{L}\left[ h_\alpha \right]  = - \frac{Z_\alpha e}{m_s v} \pd{\gyav{\chi}}{t} \pd{F_{\alpha 0}}{v} - \mathbf{v}_\chi \cdot \nabla r \pd{F_{0\alpha}}{r},
\end{equation}
where $\mathcal{L}$ is an operator that depends upon the fluctuating potential $\chi$, and represents the left-hand side of the gyrokinetic equation. Note that if alpha particles are trace (that is, $h_\alpha$ contributes negligibly in the equations for $\chi$), then the differential operator $\mathcal{L}$ is \emph{linear}, in contrast to species which are not trace. If we invert Eq.\  \eqref{schematicGK} and insert $h_\alpha$ into Eq.\ \eqref{rfluxdef}, we find that the flux can be rigorously decomposed as:
\begin{equation} \label{rfluxdiff}
\Gamma_r = - D_{rr} \pd{F_{0\alpha}}{r} - D_{r v} \pd{F_{0\alpha}}{v}.
\end{equation}
Similarly, the flux in velocity can be expressed as:
\begin{equation} \label{vfluxdiff}
\Gamma_v = -v \nu_s F_{0\alpha} - \left(\frac{1}{2} v^2 \nu_\| +  D_{v v}\right) \pd{F_{0\alpha}}{v}- D_{v r} \pd{F_{0\alpha}}{r},
\end{equation}
where the diffusion coefficients $D_{rr}$, $D_{r v}$, $D_{v r}$, and $D_{vv}$ have been introduced. 
%With the expressions \eqref{rfluxdiff} and \eqref{vfluxdiff}, valid for a trace species, the transport equation \eqref{transporteqn} becomes a linear second-order two-dimensional PDE, which can be solved with traditional methods. 
%In other words, the trace approximation means that $\chi$ can be considered a known function of space and time in the gyrokinetic equation for the alpha particle dynamics, so the ``nonlinear'' gyrokinetic equation for alphas becomes linear in the unknown $h_\alpha$, and this linearity enables us to write Eqs.~\eqref{rfluxdiff} and \eqref{vfluxdiff}. 
A new code \textsc{T3CORE}~\cite{t3core} couples to existing \textsc{GS2}~\cite{Kotschenreuther1995,Dorland2000} simulation output and uses a finite volume method to solve Eq\ \eqref{transporteqn}. With specialized diagnostics, \textsc{GS2} is used to solve for the diffusion coefficients in Eqs.\  \eqref{rfluxdiff} and \eqref{vfluxdiff} by including two trace species with the same mass and charge as alpha particles, but with different $\partial F_0/\partial r$ and/or $\partial F_0 / \partial v$. Note these ``test'' species can even be Maxwellian, and that is enough to determine the diffusion coefficients (see the proof of principle in ~\citep{Wilkie2015}). 

%where the diffusion coefficients are:
%\begin{equation} \label{diffedef}
%D_E \equiv \left\langle \sum\limits_{\sigma_\|} \int \mathcal{L}_\phi^{-1} \left[ \pd{\gyav{\phi}}{t} \right] \left\langle \mathbf{v_E} \right\rangle_{\mathbf{R}_\alpha} \cdot\nabla\psi \, \frac{Z_\alpha e  \pi B \mathrm{d} \lambda }{ \sqrt{1 - \lambda B}} \right\rangle_{t,\psi},
%\end{equation}
%Given the diffusion coefficients $D_{\psi\psi}$, $D_{\psi v}$, $D_{v \psi}$, and $D_{vv}$, we can construct the fluxes $\Gamma_\psi$ and $\Gamma_v$ for \emph{any} equilibrium distribution of alpha particles by applying \eqref{rfluxdiff} and \eqref{vfluxdiff}. A proof of principle for this kind of reconstruction was shown in reference~\cite{Wilkie2015}. 

%If we insert Eqs.\  \eqref{rfluxdiff} and \eqref{vfluxdiff} into Eq.\  \eqref{transporteqn} and take the steady-state, we arrive at a two-dimensional partial differential Eq.\  for $F_{0\alpha}(v,\psi)$. The appropriate boundary conditions are: 
%\begin{itemize}
%\item $ \Gamma_v\left(v=0,\psi\right) = 0$
%\item $F_{0\alpha} \left( v=v_{\mathrm{max}}, \psi \right) = 0$, with a suitably large $v_{\mathrm{max}}$
%\item $F_{0\alpha} \left( v, \psi = \psi_2 \right)$ given as a Dirichlet condition at the outermost flux surface $\psi_2$
%\item $ \Gamma_\psi\left(v,\psi_1\right)$ given as at the innermost flux surface $\psi_1$
%\end{itemize}
%%Equation (\ref{transporteqn}), together with Eq.\ \eqref{rfluxdiff} and the boundary conditions above, is solved with the code \texttt{T3CORE}, which requires a set of completed \texttt{GS2} flux tube turbulence simulations.

In this letter, we focus on the transport of alpha particles in gyrokinetic microturbulence. However, the novel technique described in this section, which requires about one minute of resources on a modern desktop, is generally applicable to the global turbulent transport of any trace impurity.

%If we insert (\ref{rfluxdiff}) into (\ref{transporteqn}), and take the steady-state, we arrive at the following two-dimensional partial differential equation for $F_{0\alpha}(v,\psi)$:
%\begin{equation}
%\label{sseqn}
%- \frac{1}{V'}\pd{}{\psi} \left[ V' D_\psi \pd{F_{0\alpha}}{\psi} + V' D_E \pd{F_{0\alpha}}{E} \right]  - C\left[F_{0\alpha}\right] = S_\alpha.
%\end{equation}

\section{Representative ITER scenario} \label{casesec}

\begin{table} 
\caption{\label{plasmaprops} Summary of basic profiles considered in this work. First three columns are from the CCFE public database \cite{Roach2008}, the last column is calculated from \textsc{GS2} simulation. The electron density is $10^{20}/\mathrm{m}^3$ for all radii, and the ion mix is half deuterium, half tritium, with $T_D = T_T = T_i$. The gyro-Bohm diffusivity is defined as $\rho_c^2 c_s/ a$, where $c_s \equiv \sqrt{T_e / m_D}$ is the sound speed of deuterium, and is the speed with respect to which $\rho_c$ is defined.  }
%\begin{tabular}{|@{}l|lllr|}
\begin{tabular}{llllr}
\hline
\hline
$r/ a$ & $T_i$(keV) & $T_e$(keV) & $\rho_D / a$ & $\chi_i / \chi_\mathrm{GB}$  \\
\hline
0.5 & 13.4 & 16.0 & 0.0021 & 0.74  \\
0.6 & 10.9 & 12.9 & 0.0018 & 2.3 \\
0.7 & 8.4 & 9.7 & 0.0016 & 5.4 \\
0.8 & 5.8 & 6.6 & 0.0013 & 8.7 \\
\hline
\hline
\end{tabular}
\end{table}

As a representative example of a D-T scenario of ITER, we used case 10010100 from the CCFE public database~\cite{Roach2008}: a \textsc{TRANSP} simulation of an ELMy H-mode~\cite{Budny2002} at 15 MA plasma current and a flat electron density profile. Using these data for the bulk plasma equilibrium, the flux tube code \textsc{GS2} was used to calculate the local turbulence properties (including the alpha particle diffusion coefficients) driven by instability of the ion temperature gradient (ITG) mode at four radii in the range $0.5 \le r / a \le 0.8$ (see note~\footnote{This annulus was chosen because ITG was found to be stable below $r < 0.5 a$ for this profile, and we wish to avoid the edge where ripple loss or direct orbit loss of alphas is appreciable. Yet, a significant fraction of the total alpha particles are produced in this region.}), where $a$ is the half-width of the separatrix. The intensity of the turbulence is characterized by the approximate ion diffusivity given by $q_i = - \chi_i n_i T_i'\left(r\right)$, where $n_i$, $T_i$ and $q_i$ are the deuterium density, temperature, and heat flux respectively. This is shown along with the basic plasma properties in Table \ref{plasmaprops} from our simulations, and is generally consistent with previous computational~\cite{Dimits2000,Estrada-Mila2006} and experimental~\cite{Groebner1986,Scott1990} results. The active species are deuterium, tritium, and electrons in these electrostatic simulations. 

Alpha particles produced in the region $0 < r / a < 0.5$ are assumed to enter the domain as Maxwellian ash at the local ion temperature. The distribution $F_0\left(r=0.8 a, v \right)$ at the outer edge of the domain is fixed to be the local slowing down distribution~\cite{Gaffey1976}, plus a population of helium ash at the local ion temperature to bring the total helium density to $n_{\mathrm{He}} = 10^{17} / \mathrm{m}^3$, approximately in agreement with the edge condition in \citep{,Budny2002}.

\section{Form of the modified alpha particle distribution and associated effects}

\begin{figure}
%\center{\includegraphics[width=3.4in]{figures/f03d.png}}
\center{\includegraphics[width=3.4in]{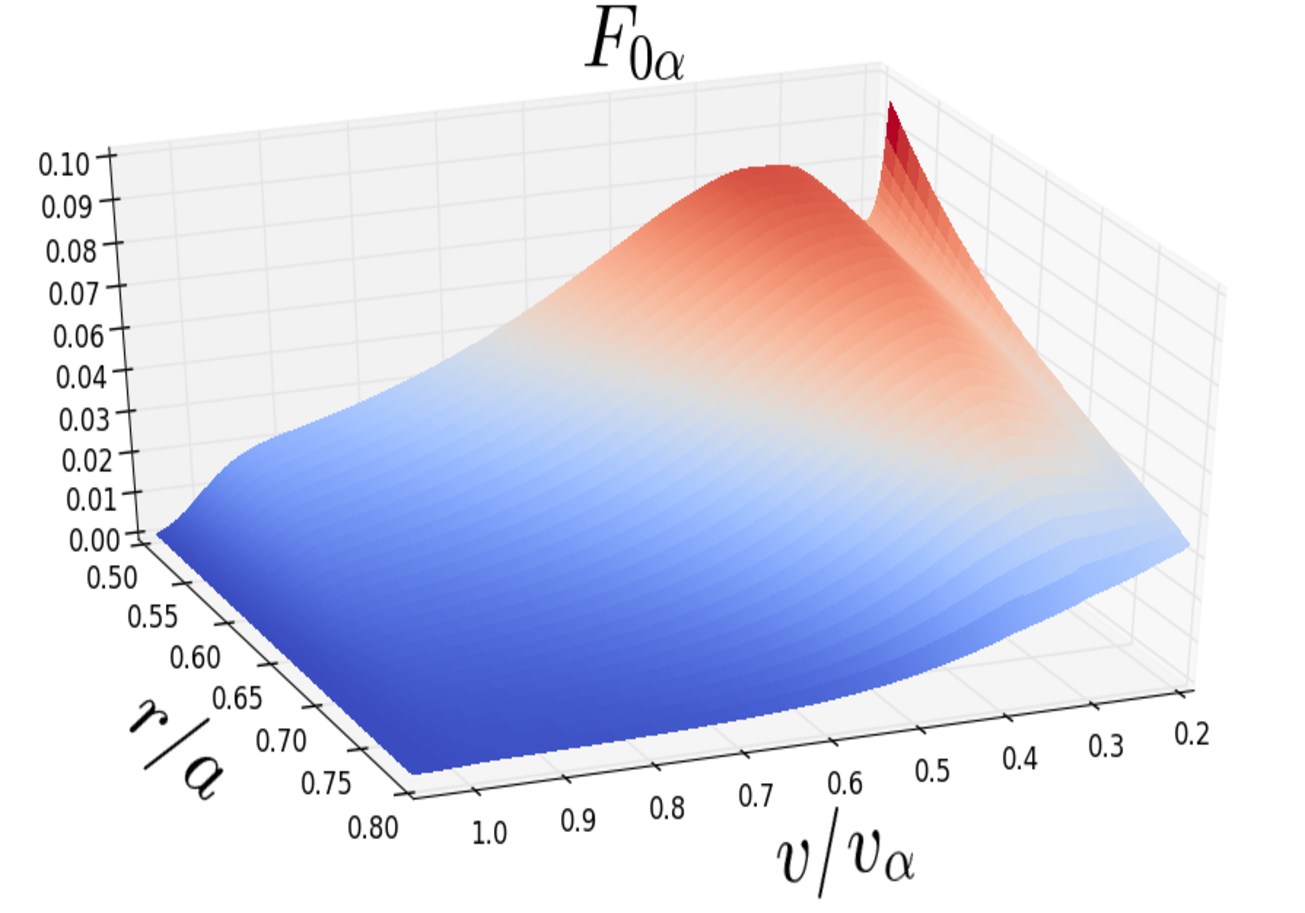}}
\caption{\label{f03d}The calculated alpha particle distribution from \textsc{T3CORE} in the presence of fusion source, turbulence, and collisions, focusing on the high-energy tail (note there exists a approximately ion-temperature Maxwellian in the region $v < 0.2 v_\alpha$. }
\end{figure}

\begin{figure}
\center{\includegraphics[width=3.4in]{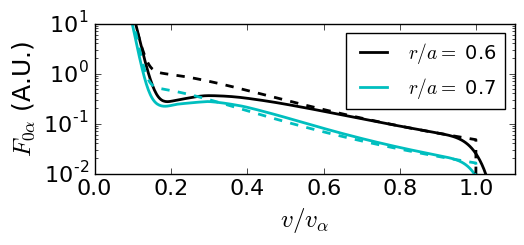}}
\caption{\label{sdcomp} Comparison at two internal radial grid points: $r = 0.6 a$ (black), and $r = 0.7 a$ (cyan) between the calculated local alpha particle distribution (solid lines) with the analytic slowing down distribution (dashed lines). A population of Maxwellian ash was artificially added to the latter so that the total helium density is the same between the two. }
\end{figure}

Using \textsc{T3CORE}, we can determine $F_{0\alpha}$, the steady-state equilibrium distribution of alpha particles in the profile described in the previous section. The resulting distribution is shown in Fig.~\ref{f03d}. A significant feature is the departure from the classical analytic slowing-down distribution~\cite{Gaffey1976} (comparison shown in Fig.~\ref{sdcomp}), including an inversion around $v/v_\alpha \sim 0.25$. Similar inversions have been seen in JET D-T experiments~\cite{JET1999}, and previous analytic transport models~\cite{Sigmar1993}. This inversion exists because the turbulent flux is a strong function of energy \cite{Hauff2009} due to finite-orbit width effects, and the transport is strong compared to collisions at suprathermal energies \cite{Wilkie2015}, ``carving out'' that part of the distribution. At high energy, the collisional slowing-down time approaches an upper bound, while the transport time continues increasing, which makes the local slowing-down distribution a better approximation. 

\begin{figure}
\includegraphics[width=3.4in]{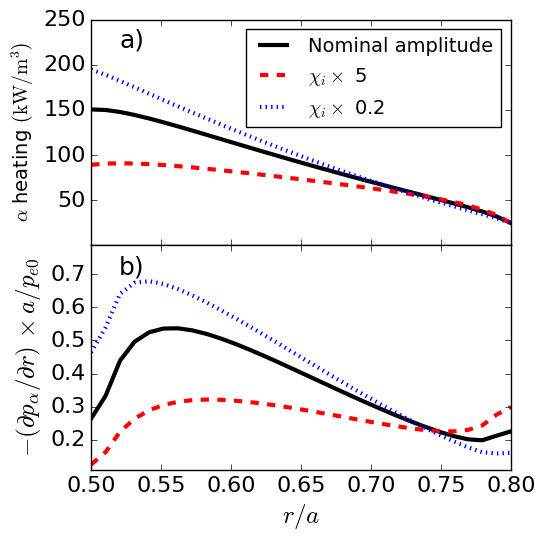}
\caption{\label{radialplots} Comparison of the radial profiles of: (a) bulk plasma heating by alpha particles, and (b) alpha particle pressure gradient normalized to the electron pressure at the magnetic axis. Displayed are cases where the nominal turbulent fluctuation amplitude and diffusion coefficients (black solid lines) are scaled up (red dashed lines) and down (blue dotted lines) by factors of five. }
\end{figure}

The modified $F_{0\alpha}$ found from simulation has an impact on several properties associated with alpha particles. Firstly, the collisional plasma heating is shown in Fig.~\ref{radialplots}~(a) to be adversely affected by the presence of turbulence due to the change in the alpha particle energy distribution. 
Furthermore, even though alpha particles have relatively low density, their pressure can account for a significant fraction of that of the total plasma. Therefore, a change in the pressure profile such as shown in Fig.~\ref{radialplots}~(b) can have a feedback effect on the magnetic geometry. 

\begin{figure}
\includegraphics[width=3.4in]{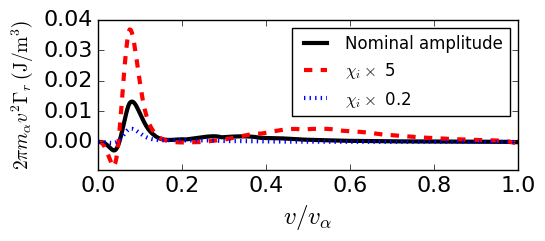}
\caption{\label{edgespectrum} Integrand of the alpha particle heat flux (which can be expressed as $q_\alpha = \int \left(m_\alpha v^2/2 \right) \Gamma_r \,4 \pi v^2 \mathrm{d}v$) at $r=0.8 a$, showing the spectrum of alpha particles exiting the domain.}
\end{figure}

If alpha particles escape the plasma at high energy, they have the potential to damage the plasma-facing components of a reactor. Therefore, an important question of alpha particle transport is if they slow down to sufficiently low energy before escaping the plasma. While we do not model the separatrix region, we can calculate the spectrum of alpha particles leaving the domain at $r=0.8a$, and this is shown in Fig.~\ref{edgespectrum}.

The intensity of microturbulence is typically quite sensitive to the gradients of density and temperature (\emph{i.e.}, ``stiff''), hence it is appropriate to examine the sensitivity of our results to the turbulence intensity in Figs.~\ref{radialplots} and \ref{edgespectrum}. There, we show the alpha particle heating, pressure, and heat flux profiles while scaling the turbulent diffusion coefficients higher and lower by a factor of five from the nominal case. There, it is clear that these key effects are sensitive to the amplitude of the turbulence.

% ??? show energy balance
% The change in the confined alpha particle energy results in a change in the energy flux of alphas at the edge of the simulation.
\section{Discussion and Conclusion}

The results presented here show that at energies around 300 keV for our ITER-like scenario, the alpha particle distribution is modified by the presence of ITG microturbulence, including an inversion that has been observed in other experiments~\cite{JET1999}.  How strong this departure is from the classical slowing-down distribution depends on the details of the turbulence. Since the slowing-down distribution is largely unaffected at high energy, we find that alpha particles largely do their job of heating the plasma, with some order unity corrections to the heating rate depending on the turbulence amplitude. Also, the alpha particle pressure profile can be significantly modified, which can in turn affect the magnetic geometry. Furthermore, the stability of Alfv\'en eigenmodes has recently been shown to be very sensitive to the alpha particle pressure profile~\cite{Rodrigues2015}, so the flattening of the alpha particle pressure profile is beneficial in this context. Our results also indicate that turbulence has only a moderate effect on the alpha particle heat flux at the high-energy part of the distribution. Only when the amplitude of the turbulence is scaled up by a factor of five does the energetic alpha particle flux become significant around 1 MeV.

Our results come with some caveats. Firstly, the trace approximation plays a central role our analysis, and this excludes some important physics. Also, the bulk equilibrium profiles were modelled from existing \textsc{TRANSP} simulations, but the need for self-consistent turbulent transport simulations is highlighted by the sensitivity of the alpha particle profiles to the turbulence amplitude. Furthermore, even though alpha particles are born isotropically in velocity space, there is the possibility that turbulence induces anisotropy in the velocity distribution, a possibility our analysis excludes.

This letter demonstrated the importance of microturbulence to the distribution of alpha particles in an ITER-like scenario and the associated consequences. The computational tool developed to solve this problem, \textsc{T3CORE}, can be coupled as a module to any of the other existing gyrokinetic or transport tools without even the need to generalize the former for non-Maxwellian distributions. With the ability to efficiently and rigorously model alpha particles in turbulence, the fusion community can make more accurate and routine predictions for the performance of ITER and devices beyond. 

\section*{Acknowledgements}

The authors would like to thank Alex Schekochihin, Michael Barnes, Greg Hammett, and Felix Parra for helpful discussions. Workshops hosted by the Wolfgang Pauli Institute in Vienna and CIEMAT in Madrid were critical to the success of this project.

This work was supported by the U.S. Department of Energy by the grants numbered DE-FG02-93ER54197 and DE-FC02-08ER54964. The nonlinear \textsc{GS2} simulations were performed on the supercomputer Hopper at NERSC.

\section*{References}

\bibliographystyle{unsrt}
\bibliography{mylib}

\end{document}